\documentclass{article}
\usepackage{spconf,amsmath,graphicx}
\usepackage{booktabs}
\usepackage{hyperref}
\usepackage[export]{adjustbox}
\usepackage{upgreek}

\title{Content Adaptive Front End For Audio Classification
}
\name{Prateek Verma \& Chris Chafe}
\address{Stanford University
\\prateekv@stanford.edu, cc@ccrma.stanford.edu}

\vspace{-2cm}
\begin{document}

\maketitle
 \begin{abstract}
We propose a learnable content adaptive front end for audio signal processing and classification. Before the modern advent of deep learning, we used fixed representation non-learnable front-ends like spectrogram or mel-spectrogram with/without neural architectures. With convolutional architectures supporting various applications such as ASR and acoustic scene understanding, a shift to a learnable front ends occurred in which both the type of basis functions and the weight were learned from scratch and optimized for the particular task of interest. With the shift to transformer-based architectures with no convolutional blocks present, a linear layer projects small waveform patches onto a small latent dimension before feeding them to a transformer architecture. In this work, we propose a way of computing a content-adaptive learnable time-frequency representation. We pass each audio signal through a bank of convolutional filters, each giving a fixed-dimensional vector. It is akin to learning a bank of finite impulse-response filterbanks and passing the input signal through the optimum filter bank depending on the content of the input signal. A content-adaptive, learnable time-frequency representation may be more broadly applicable beyond the experiments in this paper.
\end{abstract}
\vspace{-0.3cm}
\section{Introduction and Related Work}
\vspace{-0.3cm}
Humans interact with a rich palette of sounds \cite{gemmeke2017audio} in a wide range of acoustic environments. Audio signal processing which we apply to our sound world has been revolutionized by neural architectures. With the advent of transformer-based architectures \cite{vaswani2017attention}, there has been a pivot on approaching almost all problems in areas such as computer vision \cite{dosovitskiy2020image}, NLP \cite{vaswani2017attention,wei2021finetuned} and audio \cite{verma2021audio,verma2021generative,dhariwal2020jukebox}, with powerful attention-based architectures. The present work touches on ways to derive audio embeddings which have supported a variety of applications such as ASR, Audio Understanding \cite{Chung2018-Speech2Vec}, \cite{verma2020framework}, conditional audio synthesis \cite{haque2019audio,skerry2018towards} as well as style, signal transformation \cite{oord2017neural, haque2018conditional, verma2018neural}. These latent vectors are used for summarizing the audio signal's content. A classification head similar to \cite{chen2020simple, wang2021multi} is used to map these vectors to actual labels. The surge in acoustic scene understanding was first started by \cite{aytar2016soundnet} via convolutional architectures directly over raw waveforms at scale. The convolutional architecture could learn directly from audio waveforms, creating an embedding that was then projected onto image embeddings. Our work is also similar to that of the CLDNN advancement that Google proposed in \cite{sainath2015learning}, in a way by learning a front end directly from audio waveform. Our approach combines this with a mixture of expert architectures drawing from\cite{jacobs1991adaptive}. Pre-trained architectures have gone mainstream and are increasingly ubiquitous for all kinds of applications, essentially becoming universal function approximators \cite{lu2021pretrained}. In our work we provide a direction intended to further improve these architectures.
\begin{figure*}
    \includegraphics[width=0.9\textwidth,height=7.5cm,right]{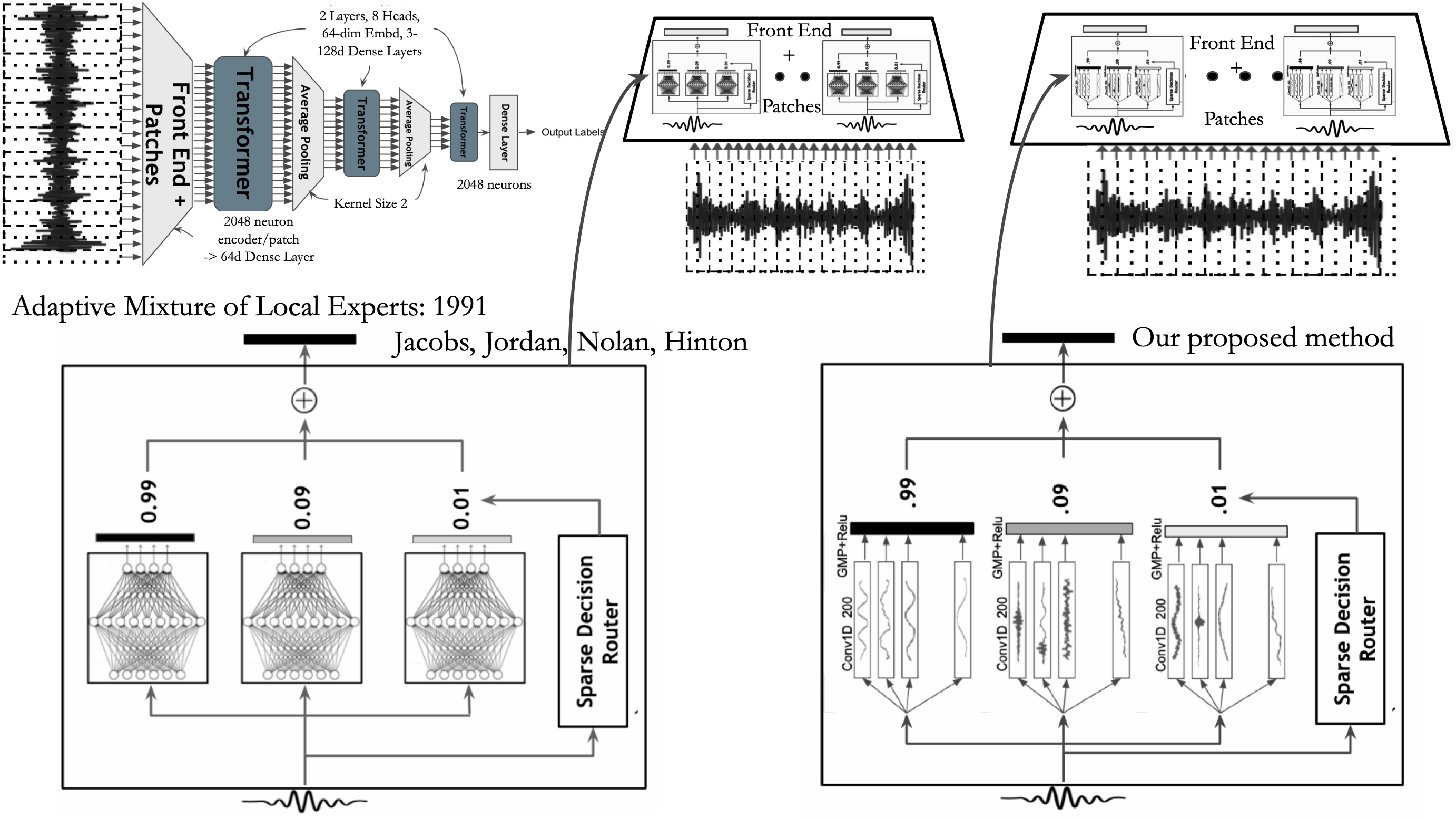}
    \caption{Our proposed method of computing the front end compared to a mixture of experts model proposed by Jacobs et. al \cite{jacobs1991adaptive}. We learn a bank of convolutional filters that can be thought of as a set of finite impulse response filterbanks.}
\end{figure*}

The contributions of this work are as follows: 1. We propose a content-adaptive front end for audio signal processing which, depending on the contents of the audio signal, \textit{routes} it to its best time-frequency representation. 2. We compare the strength of the approach to a previous architecture by substituting  it in place of the learned front end of the previous architecture, which shows a significant performance gain. 3. We show how the transformation function is both learnable and parameterized by the input signal itself. This work can be used wherever there is a learnable time-freq representation.
\vspace{-0.35cm}
\section{Methodology}
\vspace{-0.2cm}This section describes the components and the architectural choice of the current work. The goal is design of an optimal front end. For example, given an input waveform of 25ms duration, we project it on to a 64-dim vector akin to a log-magnitude mel-spectogram slice. This vector captures the contents of the input signal using a learned basis function optimized for a particular task at hand. We start with an Audio Transformer \cite{verma2021audio} as the backbone. For all our experiments, we use \cite{fonseca2020fsd50k}, which contains approximately 51k audio files labeled manually for 200 categories of sounds, with the ontology drawn from AudioSet\cite{gemmeke2017audio}. We ask the reader to refer to \cite{fonseca2020fsd50k} for a discussion of the advantages of the  \cite{fonseca2020fsd50k} over AudioSet. Having a balanced reference dataset that is freely available and a uniform way to report results are the primary reasons for this choice. We resample all the clips to have 16kHz resolution, and each 1s audio chunk we use for training inherits the label of the entire audio clip. For the second experiment, showcasing the clustering of the musical instrument families, we use the NSynth dataset \cite{engel2017neural}, comprising ten instruments and vocal sounds.
\vspace{-0.35cm}
\subsection{Our proposed front end vs. Mixture of Experts}
\vspace{-0.2cm}
A mixture of experts model was first proposed by \cite{jacobs1991adaptive} to use "experts" to learn different neural architectures for different categories of inputs. This block was extended recently, giving significant gains with transformer-based architectures at every layer within the transformer block \cite{shazeer2017outrageously}. We proposed in this work to use a mixture of expert blocks in the front end and to connect it with ideas grounded in core signal processing \cite{sainath2015learning}. We proceed in a similar way with previous architectures which feed raw waveforms to the transformer blocks \cite{verma2021audio}, \cite{akbari2021vatt}, with our approach closely following \cite{verma2021audio}. We chunk up audio signals into patches (non-overlapping) of 25ms without windowing and pass it through a linear encoder to feed it to the transformer architectures as our baseline system. This transform consists of feeding 2048 neurons directly the input of dimension 25ms (400 samples) followed by 64-dim to make it analogous to taking a 64-dim mel-spectrogram representation. This front end has also been used for pitch estimation \cite{verma2016frequency}, giving strong results and discovering comb-filters \cite{gonzalez2011pitch}. The mixture-of-experts architecture and our architecture try to develop the best architecture which will project a 25ms audio waveform onto a 64-dim vector which is then fed to the transformer architecture. We use a 6-layer transformer architecture the same as \cite{verma2021audio}, with almost the same experimental setup. The front end model, similar to \cite{jacobs1991adaptive}, consists of a given number of filter banks (experts). We experimented with banks of 2, 5, 10. Each is 2048 neurons followed by a 64-neuron layer in order to be consistent with previous work. For the case of our proposed work, we use 64-convolutional filters with a receptive field size of 320 (20ms). We again use a bank of these convolutional filters with the hyper-parameter number of filter-banks $N_{F}$, being 2,5,10. We zero-pad the input signal so that the output waveform after passing through the convolution is the same dimension as input. Thus at this layer, for each filter-bank's output from 1 till $N_F$, we get 64 x 400-dimensional output. To get a 64-dim vector from every filterbank, we use 1) average pooling over the entire 400 dimensions and 2) our proposed maximum value across 400 dimensions. This operation is followed by a relu non-linearity. We find that taking the maximum value gives a much-improved performance. This is because it tries to find an exact match with the audio waveform of interest instead of taking the average over the 25ms window.
\vspace{-0.35cm}
\subsection{Enforcing Sparsity} 
\vspace{-0.2cm}
There exist several ways of enforcing sparsity. We want to route the signals to their optimum filter bank in the most sparse way possible, i.e., each signal should always have output only from one filter bank. To achieve this routing, we have a sparse router. The current work is a  3-layer fully-connected 2048 neurons followed by a bottleneck layer with the number of neurons equal to that of the number of filterbanks. We use a softmax function to make it probabilistic but do not tune the temperature to sparsify the input. Let the output of the sparse router, which takes input as a 25ms audio patch, be $x_{sr}$. Then, after \textit{sparsifying} it, we get $x_w$, which are defined as, 
\vspace{-0.2cm}
\begin{equation}
    x_w = softmax(\alpha * softmax(x_{sr}))
\end{equation}
We opt for a significant value of $\alpha$ that does not result in overflows to ensure we almost always get sparse outputs. Several methods exist to convert continuous valued embeddings to discrete values like \cite{jang2016categorical}. However, this work aimed to make the output probabilistic while retaining a high degree of sparsity. There can be other ways to implement the sparsity block both in terms of structure and how to make the output sparse, with some of them being adjusting the temperature of the softmax output or using Gumbel-softmax \cite{jang2016categorical}.
\vspace{-0.35cm}
\subsection{Connecting the two + Loss Function}
\vspace{-0.2cm}
We multiply the output of the values of the probabilistic sparse router with that of each of the 64-dim vectors. By enforcing the output, to be as sparse as possible, we typically use the output of only one of the filterbanks from the possible $N_f$. All of it is learnable end-to-end, which means we learn an optimum filter bank and how to use it and adapt it according to the contents of the input signal. The whole architecture is optimized similarly to \cite{verma2021audio}, with the same training recipes, with the final architecture yielding a 200-dim vector that is compared with the ground truth labels with Huber Loss as the error criterion minimized. All architectures are trained for 400 epochs starting from 2e-4 till 1e-6.
\vspace{-0.35cm}
\section{Experiments, Results, and Discussion}
\vspace{-0.2cm}
This section details our experimentation. We poke into these architectures in several ways to qualitatively showcase our proposed method. We describe the proposed front end, see how the mixture weights that route the signals get clustered together and then present results with a typical acoustic scene understanding dataset. 
\vspace{-0.35cm}
\subsection{Interpreting the front-end filters} 
\vspace{-0.2cm}
For the trained model with the number of learnable filterbank being 2 for the entire FSD-50K dataset, we graph what the filters look like. These are raw weights, 64 in each filter bank for a total of 2 filter banks that we show for illustration. We see that the learned filters exhibit quite a rich behavior and that they are distinct and interpretable. The first filter bank learns low-frequency components with high-frequency modulations present in them primarily. We can see a rich structure here compared to fixed sinusoidal or Gabor wavelet functions \cite{zeghidour2021leaf}. We can see pure-sinusoids with added noises, smooth envelopes, modulated envelopes, etc. These functions can understand not only the low-frequency components of audio signals but also their modulations. As seen from the work described in \cite{verma2021audio} and \cite{verma2016frequency}, the same front end learns different basis functions/time-frequency depending on the task of interest. We show the benefits of using a rich representation which has been adaptively optimized to the task, as opposed to a fixed basis function. Further, we see from Fig. 4 that for the learnable front-end the type of mixture of expert matters. Compared to a single learnable layer/mixture-of-experts front end, the proposed
bank of learnable convolutional filterbanks (combined by the sparse connections) gives better
performance.
\begin{figure}
    \includegraphics[width=0.5\textwidth, height= 9.1cm,left]{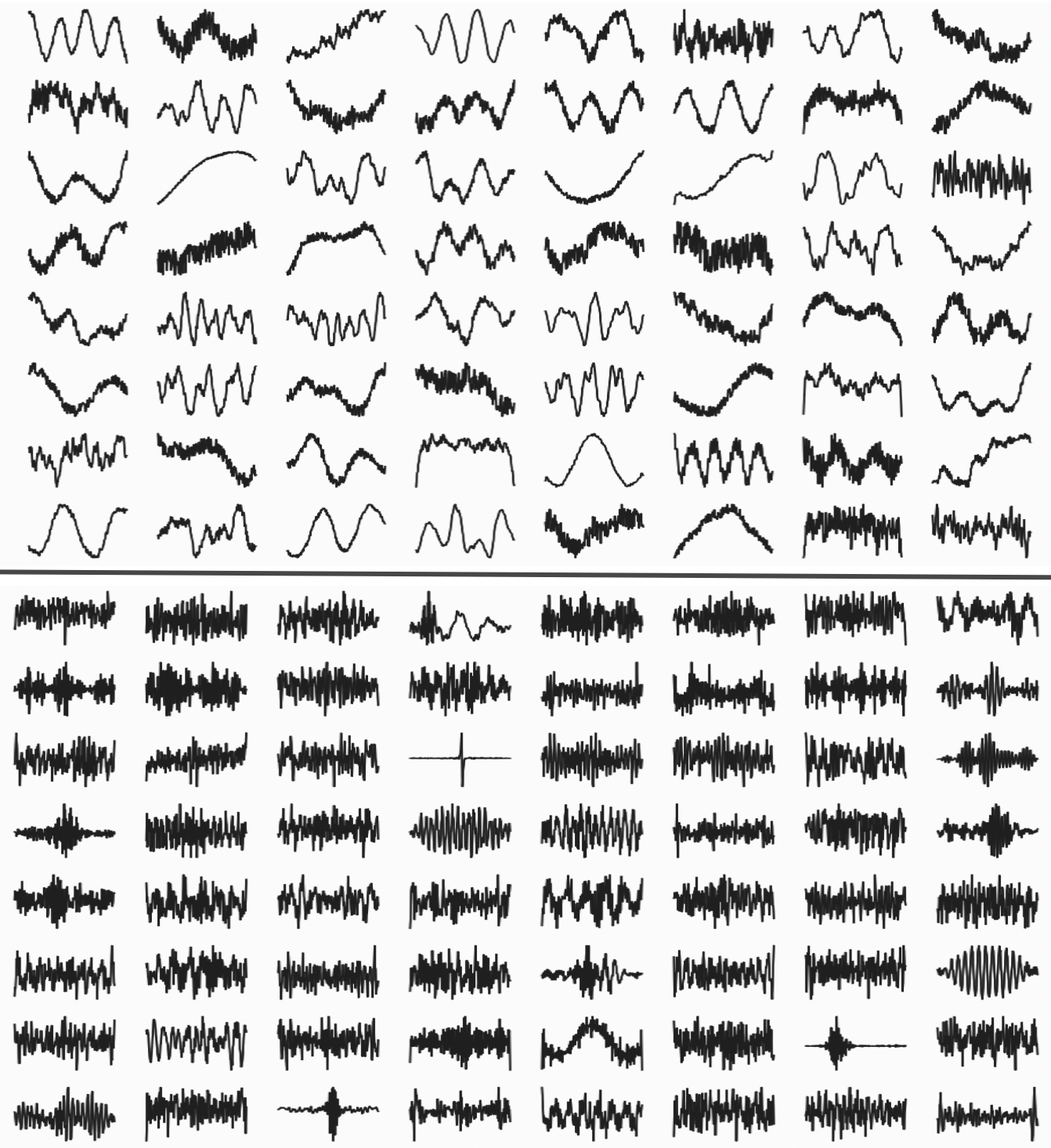}
    \caption{Conv filters from 1st/2nd filter-bank when $N_f$ = 2}
\end{figure}
\vspace{-0.35cm}
\subsection{Interpreting Mixture weights}
 \vspace{-0.2cm}
To show that the mixture weights are interpretable, we run a small experiment on the NSynth dataset \cite{engel2017neural}. The best results were obtained with a bank of 5 filterbanks. Due to rapid advancement in supervised \cite{akbari2021vatt} and unsupervised methods \cite{verma2020framework}, instrument identification with 11 categories is a relatively easy task for modern AI architectures. Once trained on the test set of instrument sounds, the mixture weights for every sound snippet is a 5-dimensional vector. The dataset consists of 11 categories: bass, brass, flute, guitar, keyboard, mallet, organ, reed, string, synth-lead, and vocal. We take for each instrument sound, the average coefficient of which filterbank (from 1-5) that it gets routed into and produce a single vector representative of the particular instrument category. This happens at 25ms resolution for each sound snippet, making optimal routing a challenging task. For each instrument category, for every sample in the validation set, we take the average of coefficients (routing which filterbank from 1-5) to get the single vector. We then compute a euclidean distance matrix between them. By just the weights, we can see a similar instrument family gets clustered together (for example, guitar, keyboard and mallet). Note this is not classification, it just learns grouping based on which filter bank is the optimum for a sound, and it groups them on its own. We also see similar instrument families being closer to each other, as seen from reed, string, and brass. Finally, the weights assigned to vocal sounds are very different from all other instrument families, which also makes sense. All these findings are just by which filter bank the signal gets routed to.
\begin{figure}
    \includegraphics[width=0.4\textwidth,height=5cm, center]{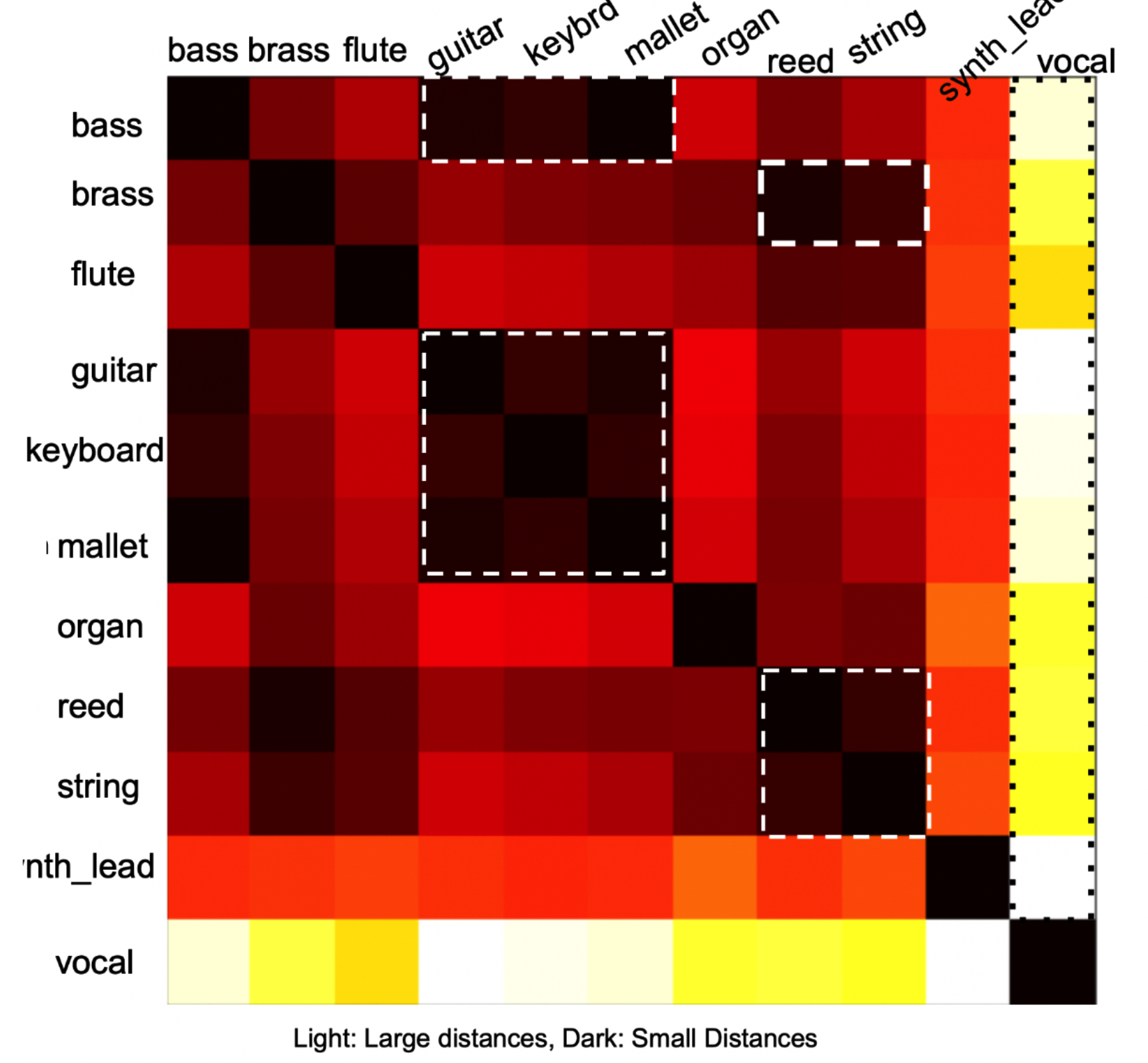}
    \caption{Distance matrix of mixture weights of a particular sound (filter-index routed to as a 5-dim vector)}
\end{figure}
\subsection{Experimental Results} 
Experiments use the FSD-50K dataset and the Audio Transformer approach  \cite{verma2021audio}. The goal is comparing the proposed front end with other enhancements such as sampling, distillation, or more data while keeping the same architecture \cite{gong2021psla}. We keep the same 6-layer Transformer with the same setting as our primary architecture. The experimental conditions are i) comparison front end same as  \cite{verma2021audio} i) mixture of experts architecture described in \cite{jacobs1991adaptive} iii) new front end with a) average pooling and b) max pooling. For the experiments, we search for the optimum number of filter banks that can be learned over 2,5,10 for routing any input signal. The sparse router for all the experiments is the same. Fig. 4 shows faster convergence and accuracy for 1s patches using mixture of expert architecture over convolutional filters. The best number of filter banks for both approaches was five on the validation set (green and brown). One possible reason might be the smaller dataset and the model over-fitting. Another likely explanation, is difficulty to make the sparse router operate as intended when we give a large number of possible routes by being able to select only one of the possible routes. Further, by using the maximum value of the output of the convolutional filter as a way of the encapsulation the content of 25ms into a 64-dim vector followed by weighing of the sparse output, we obtain almost a 4 \% increase over baseline front end as seen from Fig. 4. (baseline single filter-bank was close to orange, with the best being pink)
Table 1 shows the results of our work. 
\begin{table}[t]
  \caption{ Comparison of Mean-Average Precision over a test set of our work with other state-of-the-art architectures}
	\centering
	\begin{tabular}{|c|c|c|}
		\hline
		Neural Model Architecture & extra data & MAP score\\\hline
		DenseNet \cite{fonseca2020fsd50k} & No & 42.5\\
		Audio-Transformer \cite{verma2021audio} & No & 53.7\\
		Knowledge Distillation \cite{choi2022temporal} & No & 54.8\\
		 PLSA \cite{gong2021psla} & Yes & 56.7 \\\hline
		Ours (Bank of Filterbanks) & No & \textbf{55.2} \\\hline
		
	\end{tabular}
	\label{tab:example}
\end{table}
We believe that the gains achieved by an adaptive content front-end may be far more than reported because the transformer architecture (a 6-layer) is powerful enough to absorb the inefficiencies of a sub-optimal front-end and still achieve strong performance. It will be interesting to separate the two, compare the results keeping other blocks fixed and untrainable. 
\begin{figure}[h]
    \includegraphics[width=0.5\textwidth,height=5.5cm,center]{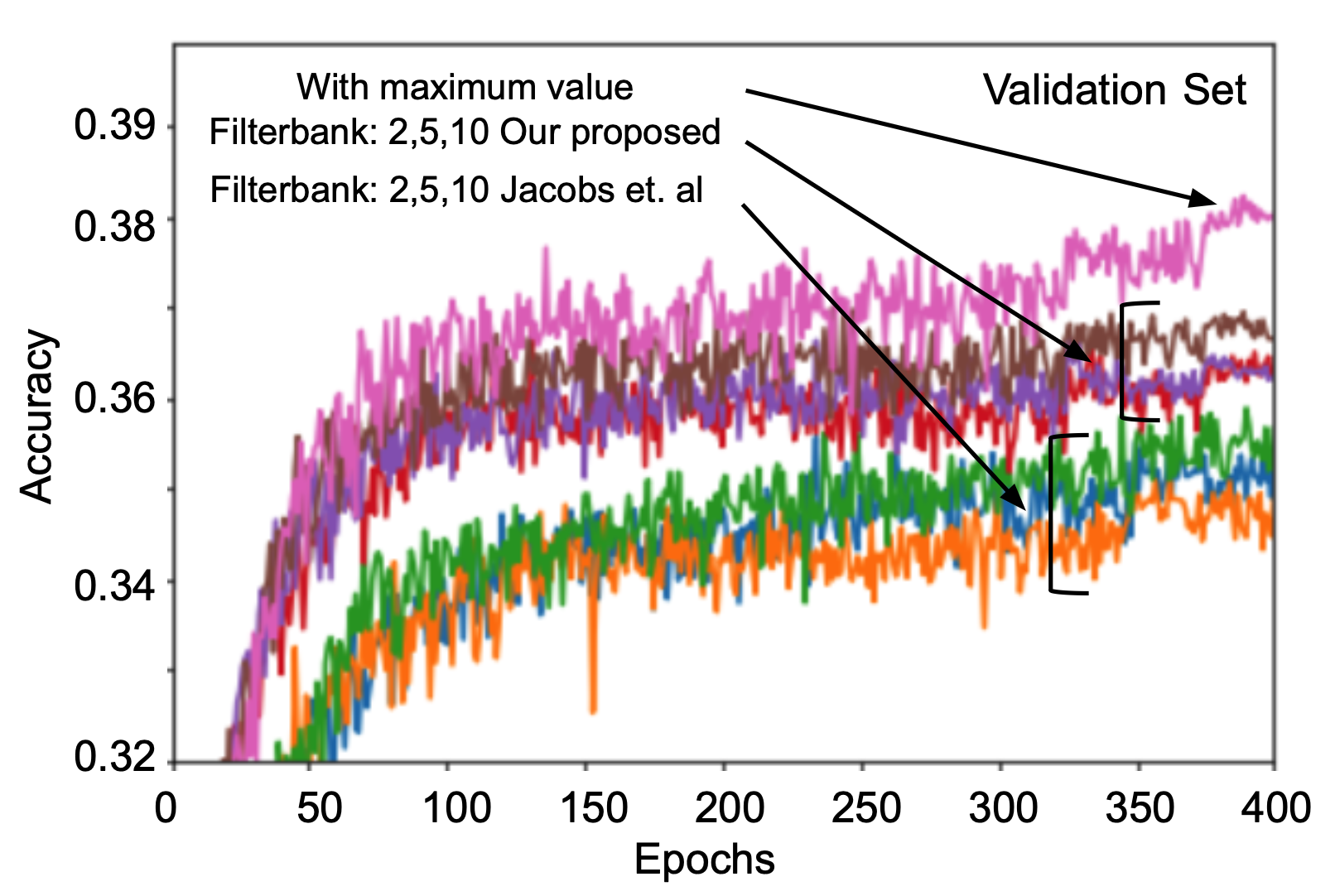}
    \caption{Top-5 accuracy(1s patches) for our bank of filterbanks (BF) front-end compared to mixture of experts (ME). Better results are obtained for BF using max-pooling(Pink) compared to the same $N_f$ =5 using avg-pooling (Brown). BF avg-pooling for $N_f$ =2 (Purple), $N_f$ =10 (Red). ME with $N_f$ = 2,10,5 (respectively, Orange, Blue, Green). The performance with Audio Transformer [5] corresponding to $N_f$ =1 is slightly below ME with $N_f$ =1 (not shown).}
\end{figure}
\section{Conclusion and Future Work} 
We showcase the strength of a content adaptive front end for acoustic scene understanding. The method produces a set of rich learned basis functions based on the task and in a way that is adaptive to the input signal. We see the front-end learning signal processing basics like onset detectors, windowing functions, first-order difference functions, pure-sinusoidal signals,and modulations being created entirely automatically resulting in a much richer toolset than pure traditional sinusoidal or Gabor functions. Improvements will be made in extending this work for various future tasks. How we compute and combine multiple, adaptive front-end time-frequency representations is an exciting new field to explore. It can potentially impact any type of audio processing that starts with raw waveforms, certainly beyond what is currently reported here.
\vspace{-0.35cm}
\section{Acknowledgement}
\vspace{-0.2cm}
This work was supported by Stanford Institute of Human-Centered AI (HAI) through a Google Cloud computing grant. 
\bibliographystyle{IEEEbib}
\bibliography{bib}
\end{document}